\documentclass[tikz,floatfix,twocolumn,superscriptaddress]{revtex4}

\usepackage{graphicx,dcolumn,bm,amsmath,amsfonts,amssymb,mathtools,subfigure,comment,tikz}
\usepackage{pgfplots,tikzscale,natbib}
\usepackage[utf8]{inputenc}
\usetikzlibrary{arrows,backgrounds,shapes,plotmarks}

\newcommand{\ket}[1]{$\left|#1\right\rangle$}
\newcommand{\Om}[1]{\small $\Omega_{#1}$}
\newcommand{\De}[1]{$\Delta_{#1}$}

\newcommand{\COMMENT}[1]{}

\begin{document}

\title{Many-body physics in two-component Bose-Einstein condensates in a cavity: fragmented superradiance and polarization}

\author{Axel U. J. Lode}
\affiliation{Wolfgang Pauli Institute c/o Faculty of Mathematics,
University of Vienna, Oskar-Morgenstern Platz 1, 1090 Vienna, Austria}
\affiliation{Vienna Center for Quantum Science and Technology,
Atominstitut, TU Wien, Stadionallee 2, 1020 Vienna, Austria}

\author{Fritz S. Diorico}
\author{Rugway Wu}
\affiliation{Vienna Center for Quantum Science and Technology,
Atominstitut, TU Wien, Stadionallee 2, 1020 Vienna, Austria}

\author{Paolo Molignini}
%\affiliation{Institute for Theoretical Physics, ETH Zurich, 8093 Zurich, Switzerland}
\author{Luca Papariello}

\author{Rui Lin}
\affiliation{Institute for Theoretical Physics, ETH Zurich, 8093 Zurich, Switzerland}

\author{Camille L\'{e}v\^{e}que}
\affiliation{Wolfgang Pauli Institute c/o Faculty of Mathematics,
University of Vienna, Oskar-Morgenstern Platz 1, 1090 Vienna, Austria}
\affiliation{Vienna Center for Quantum Science and Technology,
Atominstitut, TU Wien, Stadionallee 2, 1020 Vienna, Austria}

\author{Lukas Exl}
\affiliation{Faculty of Mathematics,
University of Vienna, Oskar-Morgenstern Platz 1, 1090 Vienna, Austria}
\affiliation{Institute for Analysis and Scientific Computing, Vienna University of Technology, Wiedner Hauptstraße 8-10, 1040 Wien, Austria}

\author{Marios C. Tsatsos}
\affiliation{S$\tilde{a}$o Carlos Institute of Physics, University of S$\tilde{a}$o Paulo, P.O. Box 369, 13560-970 S$\tilde{a}$o Carlos, S$\tilde{a}$o Paulo, Brazil.}

\author{R. Chitra}
\affiliation{Institute for Theoretical Physics, ETH Zurich, 8093 Zurich, Switzerland}

\author{Norbert J. Mauser}
\affiliation{Wolfgang Pauli Institute c/o Faculty of Mathematics,
University of Vienna, Oskar-Morgenstern Platz 1, 1090 Vienna, Austria}

\begin{abstract}
We consider laser-pumped one-dimensional two-component bosons in a parabolic trap embedded in a high-finesse optical cavity. Above a threshold pump power, the photons that populate the cavity modify the effective atom trap and mediate a coupling between the two components of the Bose-Einstein condensate. We calculate the ground state of the laser-pumped system and find different stages of self-organization depending on the power of the laser. The modified potential and the laser-mediated coupling between the atomic components give rise to rich many-body physics: 
an increase of the pump power triggers a self-organization of the atoms while an even larger pump power causes correlations between the self-organized atoms -- the BEC becomes fragmented and the reduced density matrix acquires multiple macroscopic eigenvalues. In this fragmented superradiant state, the atoms can no longer be described as two-level systems and the mapping of the system to the Dicke model breaks down.
\end{abstract}

\maketitle
\section{Introduction}
Due to the rapid advancement of optics, electronics and optoelectronic devices, the field of cavity quantum electrodynamics has also received a growing amount of interest. From scalable quantum computers \cite{Ekert2000,Kim2017,Deng2014,Schoelkopf2016} to controlling atoms in an ultracold atomic ensemble \cite{Ritsch2010,Henkel2005}: atoms coupled to light in cavities have opened new fields of promising research.
The experimental manipulation and control of ultracold atoms has drastically improved over the past decades. Considerable efforts have been put into the realization of cavity quantum electrodynamics (cavity QED) with trapped ultracold atoms in an optical cavity \cite{Brennecke2007}. The Dicke phase transition has been demonstrated in Ref.~\cite{Baumann2010} and various other setups \cite{Garraway1137,Ritsch2016,Kristian2011} have simulated the Dicke model~\cite{Dicke1954} since then. 

The impact of realizing such models goes beyond the field of ultracold atoms. Namely, cavity QED provides hybrid quantum systems~\cite{CQEDbook,Schoelkopf2008,Kurizki2015,Hannes2017,Hannes2017_2} that are one route to storing quantum information with long decoherence times. Indeed, ultracold gases are readily accessible and can have decoherence times of several seconds \cite{Deutsch2010,Schoelkopf2008}; thus ultracold atoms and, in particular, their hyperfine states (usually the ``clock states'') have bright prospects for quantum information storage. Attempts to collectively couple the microwave hyperfine ground state of an ultracold atomic ensemble in a cavity QED setting with a superconducting resonator are already being pursued by numerous groups across the world 
despite the substantial technical challenges~\cite{Verdu09,Minniberger2014,Jessen2014,Siercke2012,Bernon2013,Hattermann2017}.

To be of use as a storage for quantum information, ultracold atomic systems will inevitably have to encode correlations and entanglement into atomic many-body states~\cite{Bloch2008:Review}. For instance, collision-induced highly entangled cluster states~\cite{Jaksch1999,Mandel2003,Briegel2001} can be used as initially prepared resource states to engineer measurement-based one-way quantum computations~\cite{Raussendorf2001,Raussendorf2002,Walther2005, Kiesel2005}.  
In this work, we aim to take a first step towards an understanding of the interplay of correlations between ultracold atoms triggered by their interparticle interactions and their coupling to a cavity. Our results are promising to open up new paths on how to experimentally control multi-component condensates through optical cavities. Such a control would provide an essential building block in the development of scalable quantum computers involving systems of ultracold atoms~\cite{Cramer2013}. 

Interestingly, cavity QED can also be used to mediate interactions between atoms within an ultracold atomic ensemble~\cite{Ritsch2016}. In the dispersive regime, a transverse pump beam is coupled to the longitudinal atomic motion of the ultracold gas. The atomic motion, in turn, may populate the cavity modes. Thus, depending on the pump power, a self-organization of the atomic density is triggered~\cite{Ritsch2016, Brennecke2008}. It is possible to resonantly control these interactions: in a spinor, two- or multi-component condensate the atomic states can be coupled with each other via an optical~\cite{Brennecke2007} or a microwave cavity~\cite{Verdu09,Hattermann2017}. 

Cavity QED with ultracold atoms is essentially described by the same Hamiltonian for both microwave and optical cavities. For the microwave case a different coupling scheme is used~\cite{Verdu09} and the photon recoil can be neglected as the wavelength of microwave radiation is much larger than the typical size of the atomic cloud. For optical cavities, however, the photon wavelength is smaller than the size of the ultracold cloud and the photon recoil is therefore not negligible: the structure of the cavity mode influences the physics of cavity QED systems with an optical cavity. In recent years cavity QED experiments with ultracold atoms in optical cavities have evolved substantially~\cite{Norcia2017,Masson2017}. This development motivates us to study optical cavities in this paper. Our model is directly applicable to microwave cavities as well. However, due to the absence of photon recoil, the emergent physics are likely to be different for microwave cavities.

Theoretically, cavity QED can be described by the Tavis-Cummings Hamiltonian or generally the Dicke model \cite{Dicke1954,Tavis1968,Tavis1969}. For the case of ultracold atoms in a cavity, the mapping to the Dicke model represents a simplification that is justified as long as the considered atomic ensemble can be appropriately described by two-level systems.
However, in a degenerate ultracold ensemble, the atoms are moving and interacting with each other and the Dicke model may thus break down~\cite{lode17,Paolo}. Furthermore, the physics can change dramatically with dimensionality: the original proposal~\cite{Verdu09} to couple ultracold atoms to a superconducting microwave cavity, for instance, neglected the interparticle interactions and thereby the interesting correlations that emerge in one-dimensional multicomponent condensates~\cite{Lode2016}.

For multicomponent BECs, the basic model is the Gross-Pitaevskii equation which is a mean-field approximation and, as such, neglects correlations \cite{Pethick,PitaSandro,Bogoliubov}. The inclusion of atom-atom scattering together with atom-cavity coupling is, however, likely to affect the correlations between the atoms. Such non-trivial correlations are beyond the realm of mean-field theories and entail many-body effects. 

A representative example of such a many-body effect is the so-called \textit{fragmentation}~\cite{Spekkens,Noizieres1982} of the BEC. Fragmentation can be quantified using the reduced one-body density matrix (RDM); if the RDM has only a single macroscopic eigenvalue the system is said to be condensed~\cite{Penrose}, while if the RDM has more than one macroscopic eigenvalue the state is said to be fragmented~\cite{Spekkens,Noizieres1982,Bader,Split,Tsatsos2017}. Fragmentation has been recently demonstrated to emerge in single-component ultracold bosons coupled to a single-mode cavity for pump powers roughly four times as large as the pump power necessary to drive the system from the normal to the superradiant phase~\cite{lode17}. Importantly, fragmentation and consequently correlations are also known to be present in spinor condensates \cite{Mueller2006,Mustecapl2003,Song2014,Ho2000,Lode2016}.

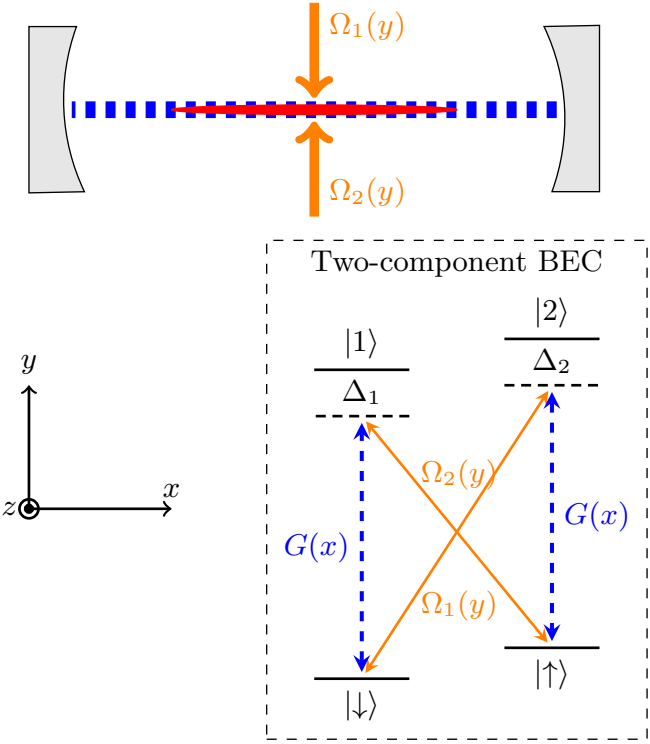
\begin{figure}
\centerline{
  %% Resize it to XXXcm wide to page width.
\resizebox{9cm}{!}{
\begin{tikzpicture}[
scale=0.5,
level/.style={thick},
virtual/.style={thick,densely dashed},
trans/.style={thick,<->,shorten >=2pt,shorten <=2pt,>=stealth},
classical/.style={thin,double,<->,shorten >=4pt,shorten <=4pt,>=stealth}]
%% Draw the energy levels.
\draw[level] (0cm,7em) -- (2cm,7em) node[midway,above] {\ket{1}};
    \draw[level] (4cm,9em) -- (6cm,9em) node[midway,above] {\ket{2}};
    \draw[level] (2cm,-13em) -- (0cm,-13em) node[midway,below] {\ket{\downarrow}};
    \draw[level] (4cm,-11em) -- (6cm,-11em) node[midway,below] {\ket{\uparrow}};
    %% Draw the virtual levels.
    \draw[virtual] (4cm,6em) -- (6cm,6em) node[midway,above] {\De{2}};
    \draw[virtual] (2cm,4em) -- (0cm,4em) node[midway,above] {\De{1}};
    %% Draw the transitions.
    \draw[trans,color=orange] (1cm,-13em) -- (5cm,6em) node[near start,right] {\Om{1}$(y)$};
    \draw[trans,color=orange] (1cm,4em) -- (5cm,-11em) node[near start,right] {\Om{2}$(y)$};
    \draw[trans,very thick,dashed,color=blue] (1cm,-13em) -- (1cm,4em) node[midway,left] {$G(x)$};
    \draw[trans,very thick,dashed,color=blue] (5cm,6em) -- (5cm,-11em) node[midway,right] {$G(x)$};
    \draw[dashed] (-1, 5) rectangle (7, -5.5) node[scale=1,shift={(-2,5)}] {Two-component BEC};  
    %% Draw x-y-z 
    \def\xaxisoffset{0}; 
    \def\yaxisoffset{0}; 
    \draw[->,thick] (-6cm+\xaxisoffset,-2em+\yaxisoffset) -- (-3cm+\xaxisoffset,-2em+\yaxisoffset) node[above] {$x$};
    \draw[->,thick] (-6cm+\xaxisoffset,-2em+\yaxisoffset) -- (-6cm+\xaxisoffset,6em+\yaxisoffset) node[above] {$y$};
    \draw[thick] (-6cm+\xaxisoffset, -2em+\yaxisoffset) circle (.2) node[left] {$z$};
    \draw[ultra thick] (-6cm+\xaxisoffset, -2em+\yaxisoffset) circle (.05); 
    %% Draw Cavity with BEC
    \def\xcavoffset{0}; % Doesn't work..
    \def\ycavoffset{0}; % Doesn't work..
    \def\xoff{-7};
    \def\yoff{5};
    \def\lensthickness{1};
    \def\lensheight{2.5};
    \path[draw, fill=gray!20] (1+\xoff+\xcavoffset,1+\yoff )--(1+\xoff+\xcavoffset,2+\yoff +\lensheight)--(1 +\lensthickness +\xoff+\xcavoffset, 2+\yoff +\lensheight) arc (160:205.5:4.5)-- cycle;
    \def\xoff{-7};
    \def\yoff{-10.525};
    \def\lensthickness{1};
    \def\lensheight{2.5};
    \path[draw, fill=gray!20,rotate=180] (1+\xoff+\xcavoffset,1+\yoff+\ycavoffset )--(1+\xoff+\xcavoffset,2+\yoff +\lensheight+\ycavoffset)--(1 +\lensthickness +\xoff+\xcavoffset, 2+\yoff +\lensheight+\ycavoffset) arc (160:205.5:4.5)-- cycle;
    \draw[-,line width=5,color=blue,dashed] (5.1+\xcavoffset,7.75+\ycavoffset) -- (-5.1+\xcavoffset,7.75+\ycavoffset);
    \draw[color=red,fill=red] (0+\xcavoffset, 7.75+\ycavoffset) ellipse (3 and 0.1);
    \draw[->,color=orange,line width=2.85] (0+\xcavoffset,10+\ycavoffset) -- (0+\xcavoffset,8+\ycavoffset) node[near start,right] {\Om{1}$(y)$};
    \draw[->,color=orange,line width=2.85] (0+\xcavoffset,5.5+\ycavoffset) -- (0,7.5+\ycavoffset) node[near start,right] {\Om{2}$(y)$};
    \end{tikzpicture}
  }
}
\caption{One dimensional quasi-condensate in a high-finesse optical cavity. The ultracold two-component bosons are pumped by two transverse pump beams that couple to the transitions between the atomic states $\vert \downarrow \rangle$ and $\vert 1 \rangle$ and between $\vert \uparrow \rangle$ and $\vert 2 \rangle$ with Rabi frequencies $\Omega_1$ and $\Omega_2$, respectively (see dashed box). The coupling to the cavity is given by $G(x)=g_0 \cos(k_c x)$ for both transitions, see text. In the self-organized phase, the cavity mode is populated (blue dashes) and acts on the atoms as a potential and effective coupling between the $\uparrow$ and $\downarrow$ component.}
\label{fig:sketch}
\end{figure}

We study the many-body physics of a one-dimensional two-component Bose-Einstein condensate in an optical high-finesse cavity as a function of the power of the applied transversal laser pumps, see Fig.~\ref{fig:sketch}. We consider a setup where the photons that populate the cavity mode through Raman scattering modify the one-body potential of the atoms and mediate a coupling between the two components of the BEC, see also Ref.~\cite{Mivehvar2017}. We go beyond Ref.~\cite{Mivehvar2017} and exploit the capabilities of the Multiconfigurational Time-Dependent Hartree Method-X (MCTDH-X) \cite{alon08,Lode2016,axel1,ultracold} to accurately~\cite{exact,exact2,exact3} include interactions and correlations between the atoms trapped by an external confinement. To this end, we extended the MCTDH-X software~\cite{ultracold} to interacting bosons with internal structure~\cite{Lode2016} that are coupled to an optical cavity. We use MCTDH-X to find the ground states and investigate the density, momentum density, the effective potential, fragmentation, cavity population, and polarization (i.e. the fraction of atoms in each component) of the ground state as a function of the strength of the pumps. 

We find that the combined system of atoms and photons undergoes two transitions. For moderate pump powers the atoms self-organize; the system is described by the Dicke model and exhibits a transition from a normal to a superradiant phase. For larger pump powers in the superradiant phase, the reduced one-body density matrix of the atoms acquires several macroscopic eigenvalues -- the BEC fragments and the combined system enters the fragmented superradiant phase~\cite{lode17}. Together with this second transition, an almost complete polarization of the atoms emerges and, simultaneously, the two-level description of the atomic ensemble -- and therewith the Dicke model -- breaks down. 
 
This paper is structured as follows. In Sec.~\ref{sec:system} we introduce the Hamiltonian, the quantities of interest and the method used. In Sec.~\ref{sec:results} we describe our results on the many-body physics of ultracold interacting two-component bosons with cavity-mediated coupling between the components.  A short discussion and outlook follow in Sec.~\ref{sec:outro}.

\section{System and Method} \label{sec:system}
The time-dependent one-dimensional many-boson Schrödinger equation in dimensionless units~\cite{units} is
\begin{equation}
 \hat{H} \vert \Psi \rangle = i \partial_t \vert \Psi \rangle. \label{TDSE}
\end{equation}
Here, $\vert \Psi \rangle$ is the many-body wavefunction of $N$ bosons in $M$ single-particle states
\begin{equation}
 \vert \Psi \rangle = \sum_{\vec{n}=(n_1,...,n_M)} C_{\vec{n}}(t) \prod_{k=1}^M \left[ \frac{\left(\hat{b}_k^\dagger\right)^{n_k}}{\sqrt{n_k!}} \right] \vert vac \rangle, \label{ansatz}
\end{equation}
where, $\hat{b}_k^{\dagger}$ is the bosonic creation operator acting on the vacuum $\vert vac \rangle$, $n_k$ is the occupation of the $k$-th single-particle state and $C_{\vec{n}}(t)$ is a time-dependent coefficient. The sum in Eq.~\eqref{ansatz} runs over all configurations $\vec{n}=(n_1,...,n_M)$  with a fixed number of particles $n_1+...+n_M=N$. 

Since the bosons have two components the single particle states $\vec{\varphi}^*_k(x;t)$ associated with the creation operators $\hat{b}_k^\dagger$ are mutually orthonormal vectors,
\begin{equation}
\vec{\varphi}^*_k(x;t) = \sum_{\xi=\uparrow,\downarrow} \phi^{\xi,*}_k (x;t)  \mathbf{1}^\xi ,
\end{equation}
made up of two functions $\phi^{\xi,*}_k$. Here $\mathbf{1}^\xi$ denotes the unit vector in the space of components. In the following we use the term ``components'' to refer to the contribution of the $\phi^\uparrow(x;t)_k$ and $\phi^\downarrow(x;t)_k$ functions to the many-body state $\vert \Psi \rangle$, while we use the terms ``orbitals'' or ``single-particle states'' to refer to the vectors $\vec{\varphi}_k(x;t)$, respectively.

The position-space Hamiltonian of the ultracold system of two-component bosons coupled to the cavity reads
\begin{equation}
 \hat{H}= \sum_{i=1}^N \hat{\underline{h}} (x_i;t) + \sum_{i<j=1}^N \hat{\underline{W}}(x_i, x_j;t). \label{H1}
\end{equation}
Here, $\hat{\underline{h}}(x;t)$ is the one-body operator that contains the kinetic energy, the confinement potential $\underline{V}(x,t)$, and a cavity-mediated term $\underline{V}^{cavity}(x,t)$, i.e.,
\begin{eqnarray}
 \hat{\underline{h}}(x;t) &=& \left[ -\frac{1}{2} \underline{\partial}^2_{x} + \underline{V}(x,t) \right] + \underline{V}^{cavity}(x,t) \label{one-body} \\
 &=& \sum_{\xi=\uparrow,\downarrow} \left[ -\frac{1}{2} \partial^2_{x} +  V_\xi (x,t) \right]  \mathbf{1}^\xi\mathbf{1}^{\xi,T} + \underline{V}^{cavity}(x,t) . \nonumber
\end{eqnarray}

The action of the cavity photons on the atoms is given by the one-body potential $\underline{V}^{cavity}(x)$:
\begin{equation}
\underline{V}^{cavity}(x)=  \left(
\begin{array}{ll}
      V^{cavity}_{\uparrow\uparrow} (x) & V^{cavity}_{\uparrow\downarrow} (x) \\
      V^{cavity}_{\downarrow\uparrow} (x) & V^{cavity}_{\downarrow\downarrow} (x) \\
\end{array}\right).
\end{equation}
The diagonal terms $V^{cavity}_{\uparrow\uparrow}$,$V^{cavity}_{\downarrow\downarrow}$ prescribe a modification of the one-body confinement $V_{\xi}$, while the off-diagonal ones, $V^{cavity}_{\downarrow\uparrow}$, $V^{cavity}_{\uparrow\downarrow}$, induce a cavity-mediated coupling between the components~\cite{Mivehvar2017}:
\begin{eqnarray}
  V^{cavity}_{\uparrow\uparrow}(x)&=& U_\uparrow \vert \alpha \vert^2 \cos^2 (k_c x),   \nonumber \\
  V^{cavity}_{\downarrow\downarrow}(x) &=& U_\downarrow \vert \alpha \vert^2 \cos^2 (k_c x) +\tilde{\delta}, \label{VCAVITY} \\
   V^{cavity}_{\uparrow\downarrow} (x) = V^{cavity}_{\downarrow\uparrow} (x) &=& \eta (\alpha + \alpha^*) \cos (k_c x). \nonumber
\end{eqnarray}
The parameters $U_{\uparrow,\downarrow}$ describe the depths of the cavity-mediated optical lattices for the two components, with $k_c$ and $\tilde{\delta}$ being the wave vector of the cavity mode and the offset between the two optical lattices, respectively. The cavity pump power $\eta$ governs the coupling between different components.

The cavity field amplitude $\alpha$ is given by the following equation of motion~\cite{lode17,Mivehvar2017}:
\begin{eqnarray}
i \partial_t \alpha(t) &=& \left[ - \Delta_c +  \sum_{k,q=1}^M \left( \rho_{kq}(t) U_{kq}(t) \right) - i \kappa \right] \alpha(t) \nonumber \\ &+& \sum_{k,q=1}^M \left(\rho_{kq}(t) \eta_{kq}(t) \right), \label{EOMc} 
\end{eqnarray}
where $\Delta_c$ is the detuning of the cavity frequency with respect to the laser pump and the decay rate $\kappa$ accounts for photons leaking out of the cavity. Here, we also introduced the matrix elements 
\begin{eqnarray}                      
U_{kq}&=& \langle \vec{\varphi}_k \vert \sum_{\xi=\uparrow,\downarrow}\mathbf{1}^\xi \mathbf{1}^{\xi,T} U_\xi \vert \alpha \vert^2 \cos^2(k_c x) \vert \vec{\varphi}_q \rangle, \label{UKQ} \\
\eta_{kq} &=& \langle \vec{\varphi}_k \vert \eta \cos(k_c x) \left(
\begin{array}{ll}
      0  & 1  \\
      1 & 0 \\
\end{array}\right)\vert \vec{\varphi}_q \rangle \label{ETAKQ} \\
&=& \eta \int dx \left( \cos(k_c x) \left[\phi^{\uparrow,*}_k (x) \phi^{\downarrow}_q (x) + \phi^{\downarrow,*}_k(x) \phi^{\uparrow}_q(x)\right]\right). \nonumber
\end{eqnarray}
The $U_{kq}$ matrix elements define the back-action of individual atomic components on the cavity field amplitude $\alpha$. The elements $\eta_{kq}$ define a coupled back-action of both atomic components on the cavity field amplitude $\alpha$ and are zero for polarized atoms. 

We note that the single distinction of the mathematical framework for the description of microwave cavities as opposed to optical cavities is the magnitude of the cavity wave vector $k_c$: for microwave cavities, the $\cos(k_c x)$ terms in equations~\eqref{VCAVITY},\eqref{UKQ},\eqref{ETAKQ} could be considered constants.

To complete our mathematical description, we consider an identical parabolic confinement for both components of the atomic cloud, 
\begin{equation}
V_\uparrow(x)\equiv V_\downarrow(x) = \frac{1}{2} x^2, 
\end{equation}
and contact interparticle interactions of atoms within the same component,
\begin{equation}
   \hat{\underline{W}}(x,x') = \sum_{\xi=\uparrow,\downarrow} \left( \mathbf{1}^\xi \mathbf{1}^{\xi,T}  \lambda^\xi_0 \delta(x-x') \right).  % + \lambda_1 \left( \sum_{\nu = 1,2,3} \mathbf{S}^{(x)}_\nu \otimes  \mathbf{S}^{(x')}_\nu \right)   \delta(x-x') 
  \label{H3}
\end{equation}
We fix the interaction strength to be weakly repulsive and slightly different for each component: for the $\uparrow$ component, we set $\lambda_0^\uparrow=0.0975$ and for the $\downarrow$ component $\lambda_0^\downarrow=0.1$ in dimensionless units; see Ref.~\cite{units} for a dimensionalized model using $^{87}$Rb atoms. Since the interactions in the $\uparrow$ component are slightly weaker, a larger population in the $\uparrow$ state is energetically favorable. Note that, for simplicity, we have neglected interparticle interactions of atoms in distinct components that are present in ultracold spinor bosons~\cite{Bruder2012,Lode2016}.

In this paper, we use the multiconfigurational time-dependent Hartree method for indistinguishable particles software~\cite{ultracold} to solve Eq.~\eqref{TDSE} for the many-body ground state of $N=100$ atoms in $M=3$ single-particle states coupled to Eq.~\eqref{EOMc} for the population of photons in the cavity. 

To compute the ground state, we propagate the coupled Eq.~\eqref{TDSE} and Eq.~\eqref{EOMc} in imaginary time to damp out all excited states. Note that the populations of the different components are varying in the process of imaginary time propagation, as the excitations of the system may have a different atom numbers in the components. The obtained ground state distributions of atoms between components are such that the total energy of the system is minimized. 

We remark that we work in dimensionless units throughout by dividing the many-body Hamiltonian by $\frac{\hbar^2}{mL^2}$ where $m$ is the mass of the considered particles and $L$ the unit of length. For example with $^{87}$Rb atoms and $L\equiv1\mu\textrm{m}$, the longitudinal extent of the system we consider is roughly $4$ to $6$ microns, yielding roughly $15$ to $20$ atoms per micron~\cite{units}. Furthermore, atomic losses are neglected in this work.

We investigate a cavity with parameters related to Esslinger's experimental setup with an optical cavity~\cite{Baumann2010} and consider a two-component system with two transversal pumps and the coupling scheme illustrated in Fig.~\ref{fig:sketch}, see also Ref.~\cite{Mivehvar2017}. The pumps and the cavity are far-red-detuned from the atomic transition.
We define the atomic and cavity detunings $\Delta_{1/2}$ and $\Delta_c$, in terms of the
energies of the involved states $\vert 1 \rangle, \vert 2 \rangle, \vert \uparrow \rangle, \vert \downarrow \rangle$, respectively, $E_{\vert 1 \rangle} = \hbar \omega_1,E_{\vert 2 \rangle} = \hbar \omega_2, E_{\vert \uparrow \rangle} = \hbar \omega_\uparrow, E_{\vert \downarrow \rangle} = \hbar \omega_\downarrow$. We fix $ E_{\vert \downarrow \rangle} \equiv 0$ and obtain the detunings: 
\begin{eqnarray*}
 \Delta_1&=&\frac{\omega_{\Omega_1}+\omega_{\Omega_2}}{2}-\omega_1,\nonumber \\
 \Delta_2&=&\omega_{\Omega_2}-\omega_{2},\nonumber \\
 \Delta_c&=&\frac{\omega_{\Omega_1}+\omega_{\Omega_2}}{2}-\omega_c.
 \end{eqnarray*}

We assume that these atomic detunings are large enough compared to the kinetic energy in the excited states $\vert 1 \rangle, \vert 2 \rangle$ such that we can eliminate them from our description, see Ref.~\cite{Mivehvar2017} for details. We consider two-photon Raman transitions to be close-to-resonant, i.e., $\omega_\uparrow \approx \omega_c - \omega_{\Omega_1} \approx \omega_{\Omega_2}-\omega_c$. The relative two-photon detuning is $\delta=\omega_\uparrow - \frac{ \omega_{\Omega_2}- \omega_{\Omega_1} }{2}$. The coupling of the $\vert \downarrow \rangle$($\vert \uparrow\rangle$)-component to the atomic excited state $\vert 1 \rangle$ ($\vert 2 \rangle$) is $G(x)=g_0 \cos(k_c x)$.  The cavity pump power is $\eta=\frac{g_0 \Omega_1}{\Delta_1}=\frac{g_0 \Omega_2}{\Delta_2}$, the cavity detuning is $\Delta_C=-42992$, the cavity loss-rate $\kappa=5555$, the $k$-vector of the cavity $k_c=4.9$, the cavity-atom coupling $U_\downarrow=\frac{g_0^2}{\Delta_1}=1, U_\uparrow=\frac{g_0^2}{\Delta_2}=2$, and the potential offset $\tilde{\delta}=\delta + \frac{\Omega_1^2}{\Delta_1} - \frac{\Omega_2^2}{\Delta_2}$ is a Stark-shifted two-photon detuning~\cite{Mivehvar2017}. 

In dimensionalized units~\cite{units}, we have ($\Delta_c$, $\kappa$, $U_\uparrow$, $U_\downarrow$)=($-2 \pi \times 4.987 \: \textrm{MHz}$, $2\pi \times 0.6444 \: \textrm{MHz}$, 
$1457.7 \: \textrm{Hz}$, $728.849 \: \textrm{Hz}$).

\section{Polarization and Fragmentation of Two-Component Bosons in a Cavity}\label{sec:results}

\begin{figure}[h]
 \includegraphics[width=1.\columnwidth]{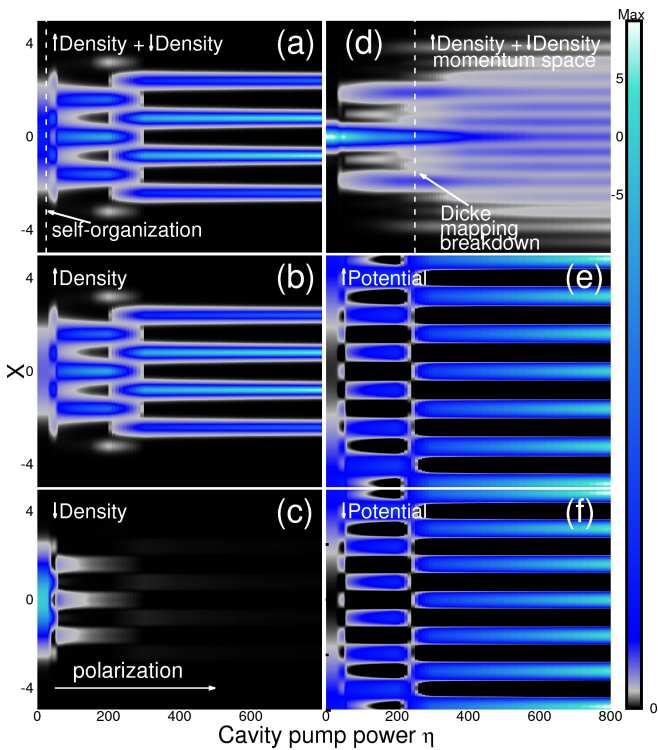}
 \caption{Tracing the self-organization of a two-component Bose-Einstein condensate in a cavity. The total [$\uparrow$ / $\downarrow$] density (a) [(b) / (c)], the total momentum density (d), and cavity-modified potential $V_{\uparrow,\downarrow}(x) + V^{cavity}_{\uparrow\uparrow,\downarrow\downarrow}(x)$ [(e) / (f)] are depicted as functions of the cavity pump power. The transition to the superradiant state in which the cavity field amplitude is nonzero and the atoms self-organize happens at $\eta_c\approx25$ [see white dashed line in panel (a)]. For larger $\eta$, the state becomes polarized  [panels (b),(c)] and the $\downarrow$ density goes to zero. The real and imaginary parts of the cavity field amplitude $\alpha$ [Fig.~\ref{fig:frag}(a)] change sign at the pump powers $\eta$ where the density (potential) changes from a two- to a three-hump and from a three- to a four-hump (-minima) structure in the superradiant phase.
 The self-organization of the two-component system results in the formation of peaks at $\pm k_c $ in the total momentum distribution. The emergence of fragmentation [cf. Fig.~\ref{fig:dens_pot}(d)] leads to the formation of additional structure with a spacing of about $k_c/3$ in the momentum distribution. See text for further discussion.}
 \label{fig:dens_pot}
 \end{figure}

We now discuss the physics of the ground state of the two-component BEC as a function of the cavity pump power. As quantities of interest, we use the reduced one-body density matrix $\bm{\rho}^{(1)}(x,x')=\langle \Psi \vert \hat{\Psi}^\dagger(x') \hat{\Psi}(x) \vert \Psi \rangle$, and its diagonal (simply called density) $\bm{\rho}(x)\equiv\bm{\rho}^{(1)}(x,x'=x)$, and the amplitude of the cavity field $\vert \alpha \vert$. Since we consider two-component bosons, the densities are also two-component quantities. The quantity $\mathbf{1}^{\uparrow} \bm{\rho}(x)$ is the \emph{component density}, as it gives the density of the $\uparrow$ component; likewise for the $\downarrow$ component. The sum of the component densities is the \emph{total} density. Figs.~\ref{fig:dens_pot}(a)--(c) show the component densities together and the total density as a function of the pump power $\eta$.

\begin{figure}
 \includegraphics[width=\columnwidth]{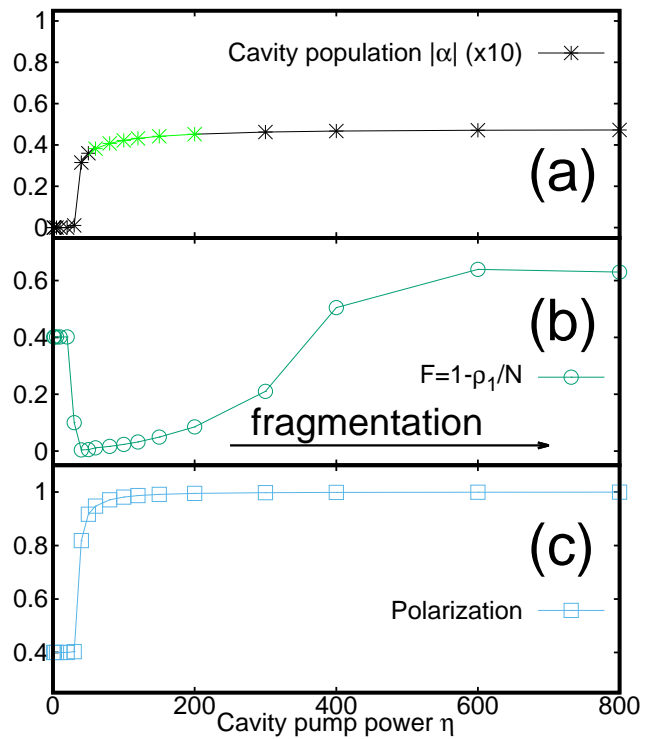}
 \caption{Cavity population, fragmentation, and polarization of two-component bosons in a cavity.
 At transition to superradiance of the Dicke model equivalent to the cold atoms in the cavity, $\eta\approx25$, the cavity field amplitude $\vert \alpha \vert$ shows a sharp increase [panel (a)]. 
 The fragmentation [panel (b)] of the state is quenched from $40\%$ to almost zero at the transition to superradiance. For larger pump powers $\eta$, concurrently with the polarization of the atoms [panel (c)], fragmentation re-emerges, however, at a smaller rate. This re-emergence of fragmentation heralds the breakdown of the mapping of the system to the Dicke model [compare momentum density in Fig.~\ref{fig:dens_pot}(d)].
 The real and imaginary parts of the cavity field amplitude $\alpha$ [panel (a)] change sign at the pump powers $\eta$ where the density changes from a two- to a three-hump and from a three- to a four-hump structure in the superradiant phase [cf. Fig.~\ref{fig:dens_pot}(a)--(c),(e)--(f)], see green/black (gray/black) part of $\vert \alpha \vert$-plot in panel (a) and text for further discussion.
 }
 \label{fig:frag}
 \end{figure}

Examining the density and its components, already reveals rich physics: close to zero pump power $\eta$ each component sees a potential that is almost harmonic as the cavity population $\alpha$ is zero [see Fig.~\ref{fig:dens_pot}(e)--(f) and Fig.~\ref{fig:frag}(a)]. The respective densities are therefore Gaussian-shaped and show no spatial modulation [Fig.~\ref{fig:dens_pot}(a)--(c)]. 
In the mapping of the system to the Dicke model, this absence of spatial modulation corresponds to the normal phase~\cite{Dicke1954,Kristian2011}. In the normal phase, the momentum distribution has a maximum at zero with no secondary peak [Fig.~\ref{fig:dens_pot}(d)]. 

As the cavity pump power crosses a threshold value of $\eta_c\approx25$, the cavity field amplitude [Fig.~\ref{fig:frag}(a)] increases and the atoms self-organize into a periodic structure as a consequence of the cavity-mediated potential [cf. Eq.~\eqref{VCAVITY}]. This self-organization is a hallmark of the transition of the system to the superradiant state of the equivalent Dicke model~\cite{Dicke1954,Kristian2011}.

With a further increase of the pump power, i.e., $\eta\in[25,120]$ the atomic ground state becomes polarized in an almost purely $\uparrow$-component state due to the cavity-mediated potential and coupling between the components [see Eq.~\eqref{VCAVITY} and Fig.~\ref{fig:dens_pot}(e),(f)]. This polarization can be quantified by the fraction of atoms in the $\uparrow$ component, 
\begin{equation*}
P=\frac{1}{N} \int dx \lbrace \mathbf{1}^\uparrow \bm{\rho}(x)\rbrace,                                                                                                                                                                                                                                                                                                                                                                                                                                                                                                                                                                                                                                                                                                                                                                                          \end{equation*}
 plotted in Fig.~\ref{fig:frag}(c). At pump powers above $\eta\approx120$, less than $1\%$ of the atoms are in the $\downarrow$ state.

The observed self-organization behavior can also be understood as a consequence of the cavity-mediated change of the one-body potential, $V_{\uparrow,\downarrow}(x) + V^{cavity}_{\uparrow\uparrow,\downarrow\downarrow}(x)$: the density of both components [cf. Fig.~\ref{fig:dens_pot}(b) and (c)] is intuitively located at the minima of the potential [cf. Fig.~\ref{fig:dens_pot}(e) and (f)].

We now turn to the emergence of correlations in the many-body state; for this purpose, we use the eigenvalues $\lbrace n_k; k=1,...,M\rbrace$ of the reduced one-body density matrix $\bm{\rho}^{(1)}(x,x')$. We quantify fragmentation by the fraction $F$ of atoms that does not correspond to the largest eigenvalue $n_1$:
\begin{equation}
 F=\frac{1}{N} \left(\sum_{k=2}^M n_k \right) = 1 - \frac{n_1}{N}.
\end{equation}
The fraction $F$ is $40\%$ at zero pump power, see Fig.~\ref{fig:frag}(b). This finding is a consequence of the slightly different interaction strengths of each component as well as the offset $\tilde{\delta}$ of the component potentials; if both interaction strengths were set equal and $\tilde{\delta}$ zero, a two-fold fragmented state with $\frac{n_1}{N}=\frac{n_2}{N}\approx50\%$ would be obtained as the ground state, because this configuration minimizes the contribution of the interactions [see Eq.~\eqref{H3}] to the total energy. 

When the atoms self-organize at $\eta_c\approx25$ and the Dicke model equivalent to the system becomes superradiant, fragmentation and $F$ are almost zero. For the range $\eta\in[25,600]$ of larger pump powers, however, fragmentation significantly increases; $F$ is larger than $0.5$ above $\eta\approx400$. The emergence of fragmentation is accompanied by a sharp growth of the atomic polarization. Above a cavity pump power of $\eta\approx120$, the system is completely polarized and almost all bosons sit in the $\uparrow$ state. The $\uparrow$ component is thus in a fragmented superradiant phase analogous to the one found for a single-component Bose-Einstein condensate in a cavity in Ref.~\cite{lode17}. This fragmented superradiant phase goes beyond the two-level physics presupposed in the Dicke model~\cite{Dicke1954} for the single-component case~\cite{lode17}.

Since the observed fragmented state in our two-component system is similar to the fragmented superradiant state found in Ref.~\cite{lode17}, it is of interest to assess the (in)applicability of the Dicke two-level picture for the present two-component case as well. For this purpose, we analyze the momentum density in Fig.~\ref{fig:dens_pot}(d). 
 
The momentum density clearly demonstrates that the Dicke model whose momentum states are at $k=\pm k_c$ and zero, qualitatively describes the physics of the system only for cavity pump powers $\eta$ for which fragmentation is essentially absent: the momentum density is essentially a three-humped structure with maxima at $k=\pm k_c$ and zero for pump powers $\eta \lesssim 250$. Here, we omitted the analysis of the component momentum densities because the ground state is almost completely polarized already for pump powers $\eta$ much smaller than $250$.

As the system enters the fragmented superradiant phase for $\eta\gtrsim250$, the Dicke model becomes inapplicable: we observe the emergence of additional structure with a $\frac{k_c}{3}$-spacing in the momentum density in our simulations in Fig.~\ref{fig:dens_pot}(d). We verified with an MCTDH-X simulation including $M=4$ orbitals that the $\frac{k_c}{3}$ spacing is not a feature of the applied approximation. Note that the momentum density corresponds to the \textit{diagonal} of the reduced momentum density matrix $\bm{\rho}^{(1)}(k,k'=k)$. This is a marked difference between the present and the fragmented superradiant state found for a single-component system in Ref.~\cite{lode17}. In the single-component case, the Dicke model also breaks down in the transition to fragmented superradiance; however, a structure with a $\frac{k_c}{2}$-spacing is formed in the \textit{off-diagonal} of the reduced momentum density matrix, $\rho^{(1)}(k,k'=-k)$ while the momentum density $\rho^{(1)}(k,k)$ is Gaussian-shaped. 
 
\section{Conclusions and Outlook}\label{sec:outro}

We have investigated the many-body physics of ultracold laser-pumped two-component bosons in a cavity. Above a first threshold of the pump power, the atoms self-organize and the system enters a superradiant state that is qualitatively described by the Dicke model. When the power of the laser pumps is increased the bosons become polarized. Above the pump power necessary for this polarization, fragmentation and correlations between the atoms emerge gradually: the reduced density matrix of the superradiant atomic ensemble acquires multiple macroscopic eigenvalues and the Bose-Einstein condensate becomes fragmented. A $\frac{k_c}{3}$-spaced pattern in the momentum distribution of the bosons heralds the breakdown of the Dicke model and the transition to a fragmented superradiant state. Our findings can be detected by a straightforward measurement of the atom numbers that populate the components and the momentum density after time-of-flight expansion.

We stress that our study explicitly includes correlations and investigates a system that is a promising candidate for ultracold-based quantum computation~\cite{Bloch2008:Review}. Understanding and possibly controlling correlations triggered in ultracold atoms interfaced with cavities enriches the field with an important contribution towards the generation of a scalable quantum computer.

We thus applied a many-body theory, the multiconfigurational time-dependent Hartree method for indistinguishable particles, and described different phases and their correlation properties; we demonstrated rich physics that result from an intricate interplay of polarization, self-organization, correlations and fragmentation.
To enable a protocol that manages the correlations of the system, further studies are needed to understand the behavior of the emergent effects as a function of the offset $\delta$ and the couplings $U_\uparrow,U_\downarrow$. Future studies may also include interparticle interactions between atoms in distinct components, multi-modal~\cite{essl2mode_1,essl2mode_2} and microwave~\cite{Verdu09,Minniberger2014} cavities, or consider more than one spatial dimension. Furthermore, the non-equilibrium dynamics of self-organization~\cite{Paolo} or the investigation of fermionic systems~\cite{Mivehar2017_2} or systems with cavity-mediated long-range interactions~\cite{luk1,luk2,Ritsch2016} are of exceptional interest.

\acknowledgments{We thank Jörg Schmiedmayer for numerous insightful discussions and comments. We acknowledge financial support by the Austrian Science Foundation (FWF) under grant No. F65 (SFB ``Complexity in PDEs''), grant No. F40 (SFB ``FOQUS''), grant No. F41 (SFB ``ViCoM''), and the Wiener Wissenschafts- und TechnologieFonds (WWTF) project No MA16-066 (``SEQUEX''). We acknowledge financial support from FAPESP, the hospitality of the Wolfgang-Pauli-Institut, computation time on the Hazel Hen cluster of the HLRS in Stuttgart and the HPC2013 cluster, financial support from the Swiss National Science Foundation and Mr. G. Anderheggen.}


\begin{thebibliography}{999}
\bibitem{units} To arrive at the dimensionless units, we divide the Hamiltonian by $\frac{\hbar^2}{mL^2}$. To arrive at the interaction parameters $\lambda_0^\uparrow=0.0975$ and 
$\lambda_0^\downarrow = 0.1$, we first fix a length scale of $L=1\mu m $. The scale of energy for the mass $m$ of $^{87}$Rb is $\hbar^2/(m L^2)=116 Hz$ and the scale of time is $mL^2/\hbar=1.37$ milliseconds. According to \cite{Olshanii}, the one-dimensional scattering lengths $\lambda$ are the three-dimensional scattering lengths $a_{3D}$ rescaled by the frequency $\omega_\perp$ of the transversal confinement by $\lambda=2Lm\omega_\perp a_{3D}/\hbar$. We pick the $\xi=\uparrow$ state and map $\lambda_0^\uparrow$ to $a_{3D}^{11}=100.4 a_0$, where $a_0$ is the Bohr radius. One obtains $\omega_\perp = \lambda_0 \hbar/(2Lma_{3D}^{11})=687.9$Hz.
\end{thebibliography}
\end{document}